


\magnification\magstep1
\parskip=\medskipamount
\hsize=6.4 truein
\baselineskip=14pt
\tolerance=500


\font\titlefont=cmbx12
\font\abstractfont=cmr10 at 10 truept
\font\authorfont=cmcsc10
\font\addressfont=cmsl10 at 10 truept
\font\smallbf=cmbx10 at 10 truept
\font\bigmath=cmsy10 scaled \magstep 4

\newdimen\itemindent \itemindent=13pt
\def\textindent#1{\parindent=\itemindent\let\par=\resetpar%
\indent\llap{#1\enspace}\ignorespaces}

\let\oldpar=\par
\def\resetpar{\oldpar\parindent=0pt\let\par=\oldpar}

\font\ninerm=cmr9 \font\ninesy=cmsy9
\font\eightrm=cmr8 \font\sixrm=cmr6
\font\eighti=cmmi8 \font\sixi=cmmi6
\font\eightsy=cmsy8 \font\sixsy=cmsy6
\font\eightbf=cmbx8 \font\sixbf=cmbx6
\font\eightit=cmti8
\def\eightpoint{\def\rm{\fam0\eightrm}
  \textfont0=\eightrm \scriptfont0=\sixrm \scriptscriptfont0=\fiverm
  \textfont1=\eighti  \scriptfont1=\sixi  \scriptscriptfont1=\fivei
  \textfont2=\eightsy \scriptfont2=\sixsy \scriptscriptfont2=\fivesy
  \textfont3=\tenex   \scriptfont3=\tenex \scriptscriptfont3=\tenex
  \textfont\itfam=\eightit  \def\it{\fam\itfam\eightit}%
  \textfont\bffam=\eightbf  \scriptfont\bffam=\sixbf
  \scriptscriptfont\bffam=\fivebf  \def\bf{\fam\bffam\eightbf}%
  \normalbaselineskip=9pt
  \setbox\strutbox=\hbox{\vrule height7pt depth2pt width0pt}%
  \let\big=\eightbig \normalbaselines\rm}
\catcode`@=11 %
\def\eightbig#1{{\hbox{$\textfont0=\ninerm\textfont2=\ninesy
  \left#1\vbox to6.5pt{}\right.\n@space$}}}
\def\vfootnote#1{\insert\footins\bgroup\eightpoint
  \interlinepenalty=\interfootnotelinepenalty
  \splittopskip=\ht\strutbox %
  \splitmaxdepth=\dp\strutbox %
  \leftskip=0pt \rightskip=0pt \spaceskip=0pt \xspaceskip=0pt
  \textindent{#1}\footstrut\futurelet\next\fo@t}
\catcode`@=12 %



\outer\def\beginsection#1\par{\vskip0pt plus.2\vsize\penalty-150
\vskip0pt plus-.2\vsize\vskip1.2truecm\vskip\parskip
\message{#1}\leftline{\bf#1}\nobreak\smallskip\noindent}


\newcount\notenumber

\def\note{\advance\notenumber by 1
\footnote{$^{\{\the \notenumber\}}$}}


\def\mapright#1{\smash{
  \mathop{\longrightarrow}\limits^{#1}}}

\def\diag{
  \def\normalbaselines{\baselineskip2pt \lineskip3pt
    \lineskiplimit3pt}
  \matrix}

\def\bigast{\mathop{\vphantom{\sum}%
                      \lower2.5pt\hbox{\bigmath\char3}}}

\def\S{\hbox{{$\Sigma$}}}

\def\P{\hbox{$\Pi$}}

\def\t{\times}
\def\pk{\pi_k(S^3)}
\def\ul{(-)^p}
\def\aut{{\rm Aut}_+^{Z_2}}
\def\trule{\hrule height 1.0 pt depth 0.5pt}
\def\Vrule{\vrule width 1.5 pt}
\def\squad{\hskip0.6em\relax}
\def\srule{\hrule height 0.1pt depth 0pt}
\def\svrule{\vrule width 0.1pt}
\def\lr{\leftrightarrow}



\rightline{Freiburg, THEP- 93/15}
\rightline{gr-qc/9308008}
\bigskip
{\baselineskip=24 truept
\titlefont
\centerline{PROPERTIES OF 3-MANIFOLDS FOR RELATIVISTS}
}

\vskip 1.1 truecm plus .3 truecm minus .2 truecm

\centerline{\authorfont Domenico Giulini\footnote*{
e-mail: giulini@sun1.ruf.uni-freiburg.de}}
\vskip 2 truemm
{\baselineskip=12truept
\addressfont
\centerline{Fakult\"at f\"ur Physik,
Universit\"at Freiburg}
\centerline{Hermann-Herder Strasse 3, D-79104 Freiburg, Germany}
}
\vskip 1.5 truecm plus .3 truecm minus .2 truecm

\centerline{\smallbf Abstract}
\vskip 1 truemm
{\baselineskip=12truept
\leftskip=3truepc
\rightskip=3truepc
\parindent=0pt

{\abstractfont
In canonical quantum gravity certain topological properties of
3-manifolds are of interest. This article gives an account of those
properties which have so far received sufficient attention, especially
those concerning the diffeomorphism groups of 3-manifolds. We give a
summary of these properties and list some old and new results
concerning them. The appendix contains a discussion of the group of
large diffeomorphisms of the $l$-handle 3-manifold.
\par}}

\beginsection{Introduction}

In the canonical formulation of general relativity and in particular
in all approaches to canonical quantum gravity it emerges that the
diffeomorphism groups of closed 3-manifolds are of particular interest.
Here one is interested in a variety of questions concerning either the
whole diffeomorphism group or certain subgroups thereof. More precisely,
one is e.g. interested in whether the 3-manifold admits orientation
reversing self-diffeomorphisms [So] (following this reference we will
call a manifold chiral, iff it admits no such diffeomorphism) or what
diffeomorphisms not connected to the identity there are in the subgroup
fixing a frame [FS][Is][So][Wi1].
Other topological invariants of this latter subgroup are also argued to
be of interest in quantum gravity [Gi1]. In the sum over histories
approach one is also interested in whether a given 3-manifold can be
considered as the spatial boundary of a spin-Lorentz 4-manifold [GH][Gi2]
(following ref. [Gi2] such 3-manifolds will be called nuclear). Motivations
for studying these questions may be found in the cited literature and
references therein.

The purpose of this paper is to present an account of results (partially
new) on these questions in a comprehensive way. A simple but non-trivial
example on which a typical diffeomorphisms group can be studied is
presented in the appendix. {\it Throughout the paper
we shall take the word 3-manifold to also imply closed,
connected and oriented unless stated otherwise}. Since 3-manifolds are not
classified, we follow the standard route by presenting the results for
some prime (defined below) 3-manifolds and indicate how the quantity in
question behaves under taking connected sums (defined below). Let us now
recall some basic facts from the subject of 3-manifolds.
A standard textbook is [He], which also contains all the relevant
references which we do not list separately.

\proclaim
Definition 1. Let $\S$, $\S_1$ and $\S_2$ be 3-manifolds.
We say that $\S$ is the connected sum of $\S_1$ and $\S_2$,
in symbols: $\S=\S_1\uplus\S_2$, if the following conditions are satisfied:
There exist two open 3-balls, $B_i\subset \S_i$, and orientation preserving
embeddings, $h_i:R_i\rightarrow\S$, where $R_i=\S_i-B_i$, such that
$h_1(R_1)\cap h_2(R_2)=h_1(\partial R_1)=h_2(\partial R_2)$ and
$h_1(R_1)\cup h_2(R_2)=\S$.\par

The operation of taking the connected sum is well defined, commutative and
associative. Taking the connected sum of any 3-manifold $\S$ with the
3-sphere results in a 3-manifold diffeomorphic to $\S$.

\proclaim
Definition 2. A 3-manifold is called a prime 3-manifold (or simply
prime), iff it is not the connected sum of two 3-manifolds
none of which is diffeomorphic to the 3-sphere.

\proclaim
Theorem 1 (Kneser). Every 3-manifold $\S$ is diffeomorphic to a finite
connected sum of prime 3-manifolds $\P_1,...,\P_n$:
$$
\S=\biguplus_{i=1}^n \P_i\,.
$$

\proclaim
Theorem 2 (Milnor). This decomposition is unique, in the sense that for
any other decomposition of $\S$ into prime 3-manifolds
$\{\P'_1,...,\P'_m\}$ it follows that $n=m$ and that there exist
orientation-preserving diffeomorphisms
$\phi_i:\P_i\rightarrow \P'_{\sigma(i)}$, $i=1,...,n$, for
some permutation $\sigma$.

Note that since the manifolds are all oriented, a prime-manifold, $\P$,
and its oppositely oriented version, $(-\P)$,  form different primes
(in the sense of Theorem 2), iff $\P_i$ is chiral.

A 3-manifold $\S$ is called {\it irreducible}, if every embedded 2-sphere
in $\S$ bounds a 3-ball. Clearly, an irreducible 3-manifold is prime.
The converse is almost true with only one exception: The ``handle''
3-manifold, $S^1\times S^2$, is the only non-irreducible prime.
Irreducibility also implies that the 3-manifold has trivial second
homotopy group. (Proof: The so called sphere-theorem
(see e.g. [He]) implies for a non-trivial second homotopy group the
existence of an element different from the identity which can be
represented by an embedded sphere. However, irreducibility enforces each
embedded sphere to be contractible.)

\beginsection{Section 1: Properties of the Diffeomorphism Group}

Associated to a manifold $\S$ we introduce the group of
$C^{\infty}-$diffeomorphisms and  several subgroups thereof.
To define these, let $\infty\in\S$ denote a fixed, preselected
point\note{The reason for this name is, that in the study of
asymptotically isolated (flat) systems, where the underlying 3-manifold
is non compact with one asymptotic region, it is convenient to
work with the one point compactification by adding a point at infinity.}.
We define:
$$
\eqalignno{
&D(\S):=\{\phi=C^{\infty}-\hbox{diffeomorphism of}\,\S\}
& (1.1a)\cr
&D_{\infty}(\S):=\{\phi\in D(\S)\,/\,\phi(\infty)=\infty\}
& (1.1b)\cr
&D_F(\S):=\{\phi\in D_{\infty}(\S)\,/\,\phi_*\vert_{\infty}=id\}\,,
& (1.1c)\cr}
$$
where $\phi_*\vert_{\infty}:T_{\infty}(\S)\rightarrow T_{\infty}(\S)$ is
the differential at $\infty$ of the map $\phi$. In words, $D$ contains all
diffeomorphisms, $D_{\infty}$ those fixing $\infty$, and $D_F$ those
which in addition fix the frames at $\infty$.
There are clearly no orientation-reversing diffeomorphisms in $D_F$ but
we take $D$ and $D_{\infty}$ to include orientation-reversing
diffeomorphisms, should they exist. Additional superscripts $+$ or $0$
may then refer to the normal subgroup of orientation-preserving
diffeomorphisms or the identity component respectively. Note also that
$D_F$ is a normal subgroup of $D_{\infty}$, but $D_{\infty}$ is no normal
subgroup of $D$.

More subgroups of $D$ may be introduced in the following way: Let
$N\subset\S$ be a closed subset. Then
$$
D(\S,relN)=\{\phi\in D(\S)\,/\,\phi\mid_N=id\}
\eqno{(1.1d)}
$$
denotes the diffeomorphisms fixing $N$ (thus generalizing (1.1b)).
If $B_{\epsilon}$ denotes any closed 3-disc neighbourhood of $\infty$
with diameter $\epsilon$ (in some fiducial metric) then clearly
$D(\S,relB_{\epsilon})\subset D_F(\S)\,\forall \epsilon$ and one may,
loosely speaking, regard $D_F(\S)$ as a limit of $D(\S,rel B_{\epsilon})$
for $\epsilon\rightarrow 0$. If one is only interested in the topological
features of the diffeomorphism groups, as we are, one may indeed replace
any $D(\S,relB_{\epsilon})$ with $D_F(\S)$, or vice versa. Now, it is in
fact $D_F(\S)$ (or, equivalently, $D(\S,rel B_{\epsilon})$) which is of
primary interst in General Realtivity (we refer to [Gi1] for a deeper
discussion of this point) and to the study of which we now turn.

The fundamental theorem,  first anounced in [RS], then elaborated
on in [HL] and [HM][McC1], aims to make precise how the
diffeomorphisms $D_F(\S)$ for any non-prime 3-manifold $\S$ are built
up from those of its prime constituents plus extra ones, and how
these extra ones can be generated by certain basic operations which
allow for more or less intuitive geometric interpretations.
In order to fully appreciate this result, we need to go through some
details and preliminary results. At the end, when grouped together,
these form what we call the theorem of Rourke de S\'a, although the
proofs of the claims in [RS], given in [HL] and [HM], do not seem to
have been given in a complete form by the authors of [RS]. In what
follows, $D$'s and $S$'s with lower case indices $i,j,k$ denote open
3-disks and 2-spheres respectively. The irreducible primes are
notationally separated from the reducible one, $S^1\times S^2$, and
will be denoted by $P_i$.

Let
$$
\S=\left(\biguplus_{i=1}^n P_i\right)\uplus\left(\biguplus_{j=1}^l
S^2\times S^1\right)
\eqno{(1.2)}
$$
be constructed in the following way: Firstly, take $P'_i=P_i-D_i$,
$1\leq i\leq n$, $\partial P'_i=S'_i$, and $l$ copies of
$I\times S^2$, labelled $I\times S^2_j$, $i\leq j\leq l$, such that
$\partial (I\times S^2_j)={S'}_{j,1}\cup {S'}_{j,2}$.
We call them the factors of $\S$. Secondly, take
a closed 3-disc, $D$, ($\partial D=S_0$) and remove  $n+2l$
mutually disjoint open 3-discs:
$$\eqalign{
B:&=D-\left\{\bigcup_{i=1}^{n}D_r\cup
\bigcup_{(j,k)=(1,1)}^{(l,2)}D_{(j,k)} \right\}\cr
\partial B&=S_0\cup\bigcup_{i=1}^n S_i\cup
\bigcup_{(j,k)=(1,1)}^{(l,2)}S_{(j,k)}\,.
\cr}
\eqno{(1.3)}
$$
We shall call $B$ the base and $S_0$ the sphere at $\infty$\note{
Since, in the notation above, it will be taken as the boundary
$\partial B_{\epsilon}$ of a neighbourhood of $\infty$.}.
Thirdly, eliminate all but $S_0$ of the boundary components
of $\partial B$ by glueing the boundaries of the factors onto
the boundary components of the base by using some identification
diffeomorphisms
$$
G_i:S'_i\rightarrow S_i\quad\hbox{and}\quad
G_{(i,k)}:S'_{(i,k)}\rightarrow S_{(i,k)}\,,
\eqno{(1.4)}
$$
such that the resulting space carries an orientation which is compatible
with the individual orientations given to the factors and the base
beforehand. For this, the maps in (1.4) must be orientation-reversing with
respect to the induced orientations. This construction is
unambiguous since any two diffeomorphisms of 2-spheres which either
preserve or reverse orientation are isotopic. Finally, we  cap-off $S_0$
with a 3-disk $D_0$ to obtain $\S$. The point $\infty$ may now be taken as
the centre of the disc $D_0$ and $D_0$ itself may be identified with one of
the $B_{\epsilon}$'s mentioned above.

Using this decomposition, we can now define four classes of diffeomorphisms
of $\S$ that will suffice to generate all orientation preserving
diffeomorphism up to isotopy (theorem 3 below). A proof may be found in
[McC1]. To unify the notation we shall commit a slight abuse of language in
not always distinguishing between $D_F(\S)$ and $D(\S,relD_0)$ or
$D_F(P_i)$ and $D(P'_i,rel S'_i)$.

1. {\it Internal diffeomorphisms:} These are diffeomorphism that
reduce to the identity when restricted to $B$. The individual
supports, $P'_i$ and $I\times S^2_i$, are disjoint in $\S$ and
elements with support in two disjoint sets of factors clearly commute.
For the handles $S^1\times S^2$ one can show that
$D(I\times S^2_i, \,rel\, \partial I\times S^2_i)$ is homotopy
equivalent to the fibre preserving diffeomorphisms not exchanging the
boundary components ($I\times S^2$ being viewed as the trivial $S^2$
bundle over $I$), and hence to $\Omega (SO(3))$, the space of based
loops of $SO(3)$. We write:
$$
D_{\rm int}=\prod_{i=1}^n D_F(P_i)\times\prod_{j=1}^l \Omega(SO(3))\,.
\eqno{(1.5)}
$$

2. {\it Exchange diffeomorphisms:} Given any two diffeomorphic
factors $P'_i$, $P'_j$ (resp. any two $I\times S^2_i$,
$I\times S^2_j$) and the associated diffeomorphism
$\phi_{ji}:\,P'_i\rightarrow P'_j$ (resp. $I\times S^2_i\rightarrow
I\times S^2_j$). Given also a diffeomorphism
$\psi_{ji}$ of $B$, exchanging $S_i$ and $S_j$ (resp. $S_{(i,k)}$ and
$S_{(j,k)}$ for $k=1,2$) in such a way, that outside some
neighbourhood in $B$ containing these but no other boundary spheres
$\psi_{ij}$ restricts to the identity.
Let it also be adjusted in such a way that it be compatible with the
glueing instructions (1.4):
$$
G_j\circ\phi_{ji}{\Big\vert}_{S'_i}=\psi_{ji}{\Big\vert}_{S_i}\circ G_i
\quad\hbox{resp.}\quad
G_{(j,k)}\circ\phi_{ji}{\Big\vert}_{S'_{(i,k)}}=
\psi_{ji}{\Big\vert}_{S_{(i,k)}}\circ G_{(i,k)}\,.
\eqno{(1.6)}
$$
Simultaneously performing $\phi_{ji}$ and $\psi_{ji}$ now defines a
diffeomorphism of $\S$ which we call an exchange of $P_i$ and $P_j$
(resp. $I\times S_i$ and $I\times S_j$). Any diffeomorphism generated
by the exchanges of the type just described is then called an exchange
diffeomorphism of $\S$.

\goodbreak

3. {\it Spin diffeomorphisms:} These are like the exchange
diffeomorphisms but concern only the two ends of handles. More
precisely, take an orientation preserving diffeomorphism
$\phi_i:\,I\times S^2_i\rightarrow I\times S^2_i$, exchanging
$S'_{(i,1)}$ and $S'_{(i,2)}$ and a diffeomorphism $\psi_i$ of $B$
exchanging $S_{(i,1)}$ and $S_{(i,2)}$ such that outside some
neighbourhood in $B$ containing these but no other boundary spheres
$\psi_i$ restricts to the identity. Let it also be adjusted in such a way
that it be compatible with the glueing instructions:
$$
G_{(i,1)}\circ\phi_i{\Big\vert}_{S'_{(i,2)}}=
\psi_i{\Big\vert}_{S_{(i,2)}}\circ G_{(i,2)}\quad\hbox{and}\quad
G_{(i,2)}\circ\phi_i{\Big\vert}_{S'_{(i,1)}}=
\psi_i{\Big\vert}_{S_{(i,1)}}\circ G_{(i,1)}\,.
\eqno{(1.7)}
$$
Performing $\phi_i$ and $\psi_i$ simultaneously defines a diffeomorphism
of $\S$ which we call a spin of the $i$'th handle. All these generate
the spin diffeomorphisms of $\S$.

\goodbreak

4. {\it Slide diffeomorphisms:} The diffeomorphisms mentioned so far
leave $B$ invariant as a set. Slides represent those elements that
mix the interior of the factors with the base $B$. Consider a fixed
factor $P'_i$ (resp. $I\times S^2_i$) and a nonintersecting
noncontractible and oriented loop
$\gamma$ in the complement of all other prime factors through $p\in B$
($\gamma$ thus represents a nontrivial element in the obvious subgroup
$\pi_1(P_i)$ (resp. $\pi_1(S^1\times S^2)$) of $\pi_1(\S)$.
Now choose $j\not =i$, cut $\gamma$ at $p$ and connect the two ends to
two different (say antipodal) points of $S_j$ so that the curve is still
nonintersecting. A thin closed
neighbourhood, $N$, of this `curve-attached-to-sphere' has the
toplogy of a solid 2-torus with an open ball removed from its
interior. $N$ has two boundary components, the two-sphere $S_j$ as
inner boundary, and a two-torus $T$ as outer boundary. An
inner collar neighbourhood $T'$ of $T$, denoted by $[0,1]\times T$
such that $1\times T=T$, can be coordinatized by
$(t,\theta,\varphi)$, where the $\varphi$ coordinate lines are
running ``parallel'' to $\gamma$ (i.e. generate $Z=\pi_1(N)$), and
$\theta$ runs along the meridians (i.e. they are contractible within
$N$). Let $\sigma:[0,1]\rightarrow [0,1]$ be a $C^{\infty}$-function,
such that $\sigma (0)=0$, $\sigma(1)=1$, and with vanishing
derivatives at $0$ and $1$ to all orders. We define the following
diffeomorphism on $T'$:
$$
s:\,(t,\theta,\varphi)\longmapsto (t',\theta', \varphi'):=
(t,\theta, \varphi+2\pi \sigma (t))\,,
\eqno{(1.8)}
$$
and continue it to the complement of $T'$ by the identity. This
defines a diffeomorphism of $\S$ with support in $T'\subset N$ which
we call a slide of $P'_j$ along $\gamma$. Analogously, instead of
the sphere $S_j$, we could have taken any of the spheres $S_{(j,k)}$
($j\not =i$ if $\gamma$ generates
$\pi_1(S^1\times S_i)$). In this case the resulting diffeomorphism is
called a slide of the $k$'th end of the $j$'th handle, $I\times
S^2_j$, along $\gamma$. We restricted attention to those $\gamma$
that where homotopically non-trivial within $\S$ since it may be
shown that slides along contractible $\gamma$ are isotopic to the
identity and that slides along the composite loop
$\gamma_1\gamma_2$ are isotopic to the composition of each individual
slide. This ends our presentation of the four classes of
diffeomorphisms.

Another important class of diffeomorphisms is given by the {\it rotations
parallel to spheres} which we now define. Given a 3-manifold $\S$,
an embedding $E:[0,1]\times S^2\rightarrow\S$, a smooth non-contractible
loop $\lambda:[0,1]\rightarrow SO(3)$ based at the identity, and a function
$\sigma$ as just defined for slides.
On $[0,1]\times S^2$ one has the diffeomorphism
$$
r:[0,1]\times S^2\rightarrow [0,1]\times S^2,\quad
(t,x)\mapsto r(t,x):=(t,(\lambda\circ\sigma(t))\cdot x)\,,
\eqno{(1.9)}
$$
through which we define a diffeomorphism of $\S$ ($Im=$ Image)
$$
R:\S\rightarrow\S,\quad p\mapsto
\cases{E\circ r\circ E^{-1}(p) & for $p\in ImE$\cr
       p & for $p\not \in ImE$\cr}
\eqno{(1.10)}
$$
which we call a {\it a rotation parallel to the spheres} $E(0)$ and
$E(1)$. Any other choice of a smooth, non-contractible loop $\lambda$
would give rise to an isotopic diffeomorphisms. If $ImE$ is a
collar neighbourhood of a sphere $S$ in $\S$, we may simply speak of
a {\it rotation parallel to $S$}.

 From (1.5) we see that a rotation parallel to a sphere $t\times S^2$
defines a diffeomorphism of the handle-manifold which is not connected
to the identity. We call it a {\it belt-twist}. Applied to a particular
factor $I\times S^2_i$ of $\S$ it is isotopic to a rotation parallel to
$S_{(i,1)}$ or $S_{(i,2)}$, which we consider as internal diffeomorphisms.
In the same way, a rotation parallel to the
connecting sphere $S_i$ of $P_i$ is considered as internal diffeomorphisms.
We call it a {\it rotation of $P_i$}. We can also define a rotation of the
$i$'th handle by a rotation parallel to a sphere enclosing $S_{(i,1)}$ and
$S_{(i,2)}$. However, this is easily seen to be isotopic to two rotations
parallel to each one of these spheres (see e.g.[Gi1]) and hence isotopic
to the identity within $D_F(\S)$.

We say that $\S$ is {\it spinorial}, iff a rotation parallel to $S_0$ --
which we simply call {\it a rotation of $\S$} -- is not connected to the
identity in $D_F(\S)$. A single handle manifold is thus not spinorial.
It is easy to see that a rotation of $\S$ is isotopic to rotating all
$P_i$'s. Thus, $\S$ is spinorial iff any of the $P_i$'s is [Gi1].

Given these definitions, we remark that there is a certain ambiguity in
the definition of slides of ends of handles or spinorial primes. To see
this, note that an alternative choice of the map $s$ in (1.8) would have
been to let also $\theta$ wind once around a full range:
$\theta'=\theta+2\pi \sigma(t)$. This would impose an additional
rotation parallel to the sphere boundary component $S_j$ of $N$ so
that the resulting diffeomorphism differs from the previous one by
the rotation parallel to $S_j$ which may be thought of as an internal
diffeomorphism and which is isotopic to the identity in
$D(P'_j,rel\, S'_j)$, iff $P_j$ is a non-spinorial prime.
In the case we slide the $k$'th end of the $j$'th handle the
resulting diffeomorphisms contains an additional rotation parallel to
the non-separating sphere $S_{(j,k)}$ ($k=1$ or $2$) which is not
isotopic to the identity within
$D(I\times S'_j,rel\,S'_{(j,1)}\cup S'_{(j,2)})$). This in fact
exhausts all ambiguities since higher rotation numbers in the
$\theta$ coordinate just result in more rotations parallel to the
spheres, which, within internal diffeomorphisms, are isotopic to the
identity for even rotation numbers. Thus we have exactly two
isotopically inequivalent definitions for sliding spinorial primes
or ends of handles. They differ by rotations of the primes or belt-twists
of the handles. Let us now state the fundamental theorems, the proofs
of which may be found in the cited literature.

\proclaim Theorem 3.
Every diffeomorphism of $\S$ is isotopic to a finite sequence of
diffeomorphisms build from the four types described above.

\proclaim Theorem 4 (Rourke de S\'a).
Given the prime factorisation (1.2), then there is a homotopy
equivalence (denoted by $\sim$; $\Omega (\cdot)$ denotes the space of
based loops in $(\cdot)$)
$$
D_F(\S)\sim\left(\prod_{i=1}^n D_F(P_i)\right)\times
                  \left(\prod_{j=1}^l \Omega SO(3)\right)
                   \times \Omega C \,.
\eqno{(1.11)}
$$

The significance of this result lies in the following: It is non-trivial
that $D_F$ is a {\it product} fibration whose fibres are the internal
diffeomorphisms. We express this by saying that internal diffeomorphisms
do not ``interact'' with external diffeomorphisms represented by
$\Omega C$. Generally, one might have expected a weaker result to hold,
namely that $D_F$ is a non-product fibration. In fact, had one
considered all diffeomorphisms and given up the restriction to those fixing
a frame, an analogous result would fail to hold. A counterexample is
given by a 3-manifold $\S$ which is the connected sum of two spinorial
primes $P_1$ and $P_2$. In $D(\S)$ the rotation parallel to the
connecting sphere $S_1$ (considered as an internal diffeomorphism)
is isotopic to the rotation parallel to the connecting sphere $S_2$.
That is, there is a path in $D(\S)$ connecting two elements in $D_F(\S)$
which (by spinoriality) cannot be connected by a path running entirely
within the internal diffeomorphisms. The fibration can thus not be a
product.

An immediate corollary of the product structure (1.11) is:
$$
\pi_k(D_F(\S))=\left(\prod_{i=1}^n \pi_k(D_F(P_i))\right)\times
           \left(\prod_{j=1}^l \pi_{k+1}(SO(3))\right)\times
           \pi_{k+1}(C)\,.
\eqno{(1.12)}
$$
Here, the only undetermined object is the space $C$. It is called the
{\it configuration space of the 3-manifold} $\S$, and, in some sense,
labels and topologizes the different relative positions of the primes
when combined to form $\S$. From theorem 3 it follows that its
fundamental group is generated by exchange-, spin-, and slide
diffeomorphisms. Furthermore, it has been shown [HM] that
$\Omega C\sim F_l^n\times \Omega C_1$ where $F_l$ is the free group
of $l$ generators and $F_l^n$ its $n$-fold product. It accounts for
the slides of the $n$ irreducible primes $P_i$ through the $l$
handles. It would certainly be desirable to continue the
factorisation to $\Omega C_1$ if possible, but generally not much seems
to be known about the detailed structure of $C_1$.  For more general
information on $C_1$ we refer to the literature [HL][HM]. In the appendix
we investigate in some more detail the group $\pi_0(D_F(\S))$ for the case
where $\S$ is given by the connected sum of $l$ handles. There we shall be
interested in learning how slides interfer with exchange- and
spin-diffeomorphisms by studying small quotient groups with obvious
representations. Heuristic arguments in canonical quantum gravity suggest
that the different representations of $\pi_0(D_F(\S))$ characterize
different sectors in quantum gravity [Is]. Usually attention is restricted
to one-dimensional representations, but this seems unnecessarily
restricitve [Gi3]. However, special proposals for the construction of
quantum states might be employed to preclude certain sectors [HW][GiL].
In theorem A of the appendix it is e.g. shown that  restriction to abelian
sectors implies that for a $l>2$ handle manifold, spins followed by
exchanges are necessarily represented trivially. This is a purely
kinematical result and independent of any requirement as to how to
construct the quantum states.
In contrast, the considerations in [HW][GiL] made essential use of the
no-boundary proposal for the construction of quantum states.

We can now introduce the properties of 3-manifolds which we wish to
give information about in this article. We indicate how these properties
behave under taking connected sums so that we can eventually restrict
attention to prime 3-manifolds. A table summarizing their properties is
then presented at the end of section 2.

$\bullet$ {\it Chirality} : A manifold is called chiral, iff it does not
allow for an orientation reversing self diffeomorphism. It follows
imediately from theorem 2 that a 3-manifold is non-chiral iff
no prime in its prime decomposition is chiral. Chirality for the relevant
spherical primes is nicely demonstrated in [Wi1]. Chirality of the flat
3-manifolds $R^3/G_i$, $i=3,4,5,6$, may be shown by inspection [McC2],
using results from [LSY]. In the table $+$ stands for chiral and $-$ for
non-chiral. Chirality is abreviated by $C$ and is listed in the fourth
column.

$\bullet$ {\it Spinoriality} : A 3-manifold is called spinorial, iff a
rotation parallel to the boundary sphere of an embedded 3-disc is not
in the connected component of $D_F$, where $D_F$ stabilizes a fixed
interior point (and all frames at this point) of the disc. It can be
shown (e.g. [Gi1]) that a 3-manifold is non-spinorial, iff no
prime in its prime decomposition is spinorial. In the table $+$ stands
for spinorial and $-$ for non-spinorial. Spinoriality is abreviated by
$S$ and is listed in the third column.

$\bullet$ {\it Nuclearity} : A 3-manifold is nuclear, iff it is the
spacelike boundary of a Lorentz 4-manifold with $SL(2,C)$ spin
structure. The necessary and sufficient condition for $\S$ to be nuclear
is that the $\{0,1\}-$ valued, so-called Kevaire semi-characteristic,
$u(\S)$, is $0$ [GH]. It can be shown that a 3-manifold is nuclear,
iff the number of nuclear primes in its prime decomposition is
odd. For disconnected $\S$ one has the following simple rule
[Gi2]: $\S$ is nuclear, iff the number of components with even number of
nuclear primes in their decomposition is even.
In the table $+$ stands for nuclear ($u=0$) and $-$ for non-nuclear
($u=1$). Nuclearity is abreviated by $N$ and is listed in the fifth column.
There, $(-)^p$ stands for: $+$ if $p$ even and $-$ if $p$ odd; a $*$
stands for: has to be decided on a case-by-case anylysis.

$\bullet$ {\it Homology groups of $\S$} : The homology groups are merely
listed for reasons of completeness. Let $A$ denote the operation of
abelianizing a group and $F$ that of taking the free part of a finitely
generated abelian group. The homology, $H_*$ (abreviating the zeroth to
third homology group as a row-quartuple), is then given in terms of the
fundamental group: $H_*=(Z,A\pi_1,FA\pi_1,Z)$.
The fundamental group of a connected sum of two 3-manifolds, $\S_1$ and
$\S_2$, is the free product (denoted by $*$)
of the individual ones: $\pi_1(\S_1\uplus \S_2)=\pi_1(\S_1)*\pi_1(\S_2)$.
It is infinite if neither of the two manifolds is simply connected.
For the homology this implies:
$H_*(\S_1\uplus \S_2)=(Z,H_1(\S_1)\times H_1(\S_2),
H_2(\S_1)\times H_2(\S_2),Z)$.
Since $H_2=FH_1$, it is enough to list $H_1$
which is done in the sixth column. There, the symbol $A\pi$ denotes the
abelianisation of the group represented by $\pi$.

$\bullet$ {\it Homotopy Groups of $D_F(\S))$} : In general relativity, the
classical configuration space, $Q$, satisfies $\pi_k(D_F)=\pi_{k+1}(Q)$
so that we obtain information on its topology by studying the topology of
$D_F$. It is explained in [Gi1]
in what sense this is equally valid for the configuration space of closed
and open universes. Theorem 4 shows how far
these groups are fixed by the corresponding ones for the primes. It tells
us that the latter ones are contained as sub- and factor groups.
The zeroth and first homotopy group of $D_F(\S)$ (resp. the first and
second homotopy group of $Q(\S)$) are listed in the seventh and eighth
column, and the higher ones $\pi_k(D_F(\S))$ for $k\geq 2$ (resp.
$\pi_{k+1}(Q(\S))$) are reduced to those of spheres of dimension two
and three in the nineth column. Calculations for $\pi_0(D_F(\P))$, where
$\P$ is a spherical prime, where first presented in [Wi1]. Details for
the other calculations are given in [Gi1]. In the seventh column
the symbols $Aut^{Z_2}_{+}(G)$ are interpreted as follows: By $Aut(G)$
(where $G$ is the fundamental group of the prime $P$, which we identify
with $\pi_1({\infty},P))$ we denote the automorphism group of $G$. It is
generated by $D_{\infty}$ via its action on the fundamental group.
$Aut_{+}(G)$ denotes the subgroup of index 2 generated by the orientation
preserving diffeomorphisms $D^+_{\infty}(P)$. Finally,
$Aut^{Z_2}_{+}(G)$ denotes a central $Z_2$ extension thereof (due to
spinoriality of $P$), where the
extending $Z_2$ is generated by a rotation parallel to a sphere whose
bounding disc contains ${\infty}$. $St(3,Z)$ is a so-called Steinberg
group (see e.g. paragraph 10 in [Mi] for more information about $St(n,Z)$)
which is a central $Z_2$-extension of $SL(3,Z)$. It is a perfect group
(i.e. its own commutator subgroup). Finally, a $*$ means: has to be
decided on a case by case analysis.

$\bullet$ {\it Validity of Hatcher Conjecture} :
For the particular class of spherical primes (explained below), the
information given about $\pi_k(D_F)$ depends on the validity of the
so-called Hatcher conjecture. This motivates to indicate its status
for the relavant primes within the list. It states that the
diffeomorphism group for a spherical prime is (as a topological space)
homotopy equivalent to the space of isometries (of the obvious metric).
A weak implication thereof (called the weak conjecture) is that these
spaces have isomorphic zeroth homotopy group. The calculation of
$\pi_0(D_F)$ only depends
on the validity of the weak conjecture, whereas those for $\pi_k(D_F)$,
$k\geq 1$, depend on the full conjecture. A $+$ indicates validity of the
full conjecture, a $w$ validity the weak form and a ? that we do not have
any information. The Hatcher conjecture is abreviated by $HC$ and is listed
for spherical primes in the second column.

\beginsection{Section 2: The Prime 3-Manifolds}

In the table presented at the end of this section we list the relevant data
for all known prime 3-manifolds except the non-sufficiently large
$K(\pi,1)$. The first column contains their conventional names as already
used in the physics literature (e.g. [Wi1]). The top line names the columns
as outlined above. Below this line, the table is divided into three
disconnected parts (framed by bold lines) which we call subtables, the
first and third of which are again subdivided by bold
lines into so-called blocks. The first subtable contains the known
prime 3-manifolds with finite fundamental group (which is clearly identical
to the set of all known 3-manifolds with finite fundamantal group).
They are all of the form $S^3/G$ where $G$ is a finite subgroup of $SO(4)$
with free action on $S^3$.
They are also called spherical primes. The first two blocks contain those
$S^3/G$ with $G$ non-cyclic, the third $G=Z_p$ for $p>2$, and the fourth
$G=Z_p$ for $p\leq 2$, i.e. the real projective 3-space, $RP^3=S^3/Z_2$,
and the 3-sphere itself.

The first block has $G\subset SU(2)$ and the resulting 3-manifolds are
homogeneous. The indexing integer has range $n\geq 1$.
In the second block $G$ is not contained in any $SU(2)$ and the manifolds
are not homogeneous. The order of $Z_p$, i.e. $p$,
is coprime to the order of the group the $Z_p$ is multiplied with and $p>1$
in the first six cases. In the remanining two cases $p=1$ is also an
allowed value. The other indexing integers have ranges $n\geq 1$,
$m\geq 2$, and $k\geq 3$.
The third block contains the so-called lens spaces. Since $Z_p$ can act in
different, non-equivalent (i.e. not conjugate by a diffeomorphism) ways,
they are labelled by $L(p,q)$ with an additional integer $q$ coprime to
$p$. Here, $q_1$ stands for $q=\pm 1$ mod $p$, $q_2$ for $q\not =\pm 1$
mod $p$ and $q^2=1$ mod $p$, $q_3$ for $q^2=-1$ mod $p$, and $q_4$ for
the remaining cases. Amongst all $L(p,q_2)$ are those of the form
$L(4n,2n-1)$, $n\geq2$.
For those the $+$ is valid in the $HC$-column and $w$ for the remaninig
cases. The $L(p,q_1)$'s are the only homogeneous lens spaces. Finally, we
note that it is a still open conjecture (involving some subconjectures)
that this list comprises all 3-manifolds of finite fundamental group
(see e.g. [Th]). Presentations of the finite fundamental groups occuring
here may be found in [O] or [Wi1].

The second subtable consists of a single member, namely the only
non-irreducible prime: $S^2\times S^3$. The third subtable
comprises the irreducible primes with infinite fundamental group
which in addition are sufficiently large (SL). They all fall into the
class of $K(\pi,1)$ spaces (Eilenberg-MacLane spaces), that is spaces
whose only non-vanishing homotopy group is the first. The condition of
being sufficiently large means that these 3-manifolds
allow an embedding of a closed Riemannian surface such that the
induced homomorphism on the fundamental groups is injective. Physically
speaking, a non-contractible loop on the embedded surface is also not
contractible within the ambient 3-manifold. In particular, the fundamental
group of a SL 3-manifolds contains as subgroup the fundamental group of a
Riemannian surface. A sufficient condition for for an irreducible manifold
to be SL is that the first Homology group (which is the abelianisation of
the fundamental group) is infinite. The reason why we restrict to the
subclass of sufficienty large ones is simply that not much seems to be
known for general $K(\pi,1)$'s. Now, the first block contains all
3-manifolds of the form $R^3/G$, where $G$ is discrete subgroup of the
affine group in 3 dimensions that acts freely and properly discontinously
on $R^3$. They comprise all flat 3-manifolds (i.e. admitting a flat
metric).
$G_1$ is equal to $Z\times Z\times Z$ (i.e. $R^3/G_1$ is just the 3-torus),
and $G_2,\dots ,G_6$ are certain extensions of the groups $Z_2$, $Z_3$,
$Z_4$, $Z_6$ and $Z_2\times Z_2$ respectively by
$G_1$ (i.e. $R^3/G_2,\dots ,R^3/G_6$ are further quotients of the 3-torus).
The second block contains manifolds of
the form $S^1\times R_g$ where $R_g$ denotes a Riemannian surface of genus
$g$. The third block represents all other sufficiently large $K(\pi,1)$
primes.
For relativists, an interesting property of $K(\pi,1)$ primes is that no
connected sum containing at least one of them admits a Riemannian
metric of everywhere positive scalar curvature. Moreover, if it admits a
nowhere negative scalar curvature metric, it must in fact be flat,
i.e. the 3-manifold must be one of the six ones listed in the first block.
This has been proven in [GL], and its significance for General Relativity
pointed out in [Wi2].
\vfill\eject




\advance\hoffset by -1.1truecm
\advance\hsize by 2.3truecm
\advance\vsize by 2truecm

\centerline{PROPERTIES OF PRIME 3-MANIFOLDS}
{\eightpoint
\setbox\strutbox=\hbox{\vrule height12truept depth7truept width0pt}
$$
\vbox{
\offinterlineskip
\halign{\strut
    \Vrule\quad\hfil $#$ \quad\hfil   & \svrule\hfil $\squad #\squad $\hfil
& \svrule\hfil $\squad #\squad $\hfil & \svrule\hfil $\squad #\squad $\hfil
& \svrule\hfil $\squad #\squad $\hfil & \svrule\squad\hfil $#$ \squad\hfil
& \svrule\squad\hfil $#$ \squad\hfil  & \svrule\hfil $\squad #\squad $\hfil
& \svrule\squad\hfil $#$ \squad\hfil    \Vrule  \cr
\noalign{\trule}
{\rm Prime\,\P}\,  & {\rm HC} & {\rm S} & {\rm C} & {\rm N} & H_1(\P) &
                     \pi_0(D_F(\P)) & \pi_1(D_F(\P)) & \pi_k(D_F(\P))
                     \cr
\noalign{\trule\vskip 6.0pt\trule}
S^3/D^*_8&+&+&+&-&Z_2\t Z_2&O^*&0&\pk \cr
\noalign{\srule}
S^3/D^*_{8n}&+&+&+&-&Z_2\t Z_2&D^*_{16n}&0&\pk \cr
\noalign{\srule}
S^3/D^*_{4(2n+1)}&+&+&+&+&Z_4&D^*_{8(2n+1)}&0&\pk \cr
\noalign{\srule}
S^3/T^*&?&+&+&-&Z_3&O^*&0&\pk  \cr
\noalign{\srule}
S^3/O^*&w&+&+&+&Z_2&O^*&0&\pk  \cr
\noalign{\srule}
S^3/I^*&?&+&+&-&0&I^*&0&\pk  \cr
\noalign{\trule}
S^3/D^*_8\t Z_p&+&+&+&-&Z_2\t Z_{2p}&Z_2\t O^*&Z&\pk\t\pk \cr
\noalign{\srule}
S^3/D^*_{8n}\t Z_p&+&+&+&-&Z_2\t Z_{2p}&Z_2\t D^*_{16n}&Z&\pk\t\pk \cr
\noalign{\srule}
S^3/D^*_{4(2n+1)}\t Z_p&+&+&+&+&Z_{4p}&Z_2\t D^*_{8(2n+1)}&Z&\pk\t\pk \cr
\noalign{\srule}
S^3/T^*\t Z_p&?&+&+&-&Z_{3p}&Z_2\t O^*&Z&\pk\t\pk  \cr
\noalign{\srule}
S^3/O^*\t Z_p&w&+&+&+&Z_{2p}&Z_2\t O^*&Z&\pk\t\pk  \cr
\noalign{\srule}
S^3/I^*\t Z_p&?&+&+&-&Z_p&Z_2\t I^*&Z&\pk\t\pk  \cr
\noalign{\srule}
S^3/D'_{2^k(2n+1)}\t Z_p&+&+&+&+&Z_p\t Z_{2^k}&Z_2\t D^*_{8(2n+1)}&Z&
                                                            \pk\t\pk  \cr
\noalign{\srule}
S^3/T'_{8\cdot 3^m}\t Z_p&?&+&+&-&Z_p\t Z_{3^m}&O^*&Z&\pk\t\pk \cr
\noalign{\trule}
L(p,q_1)&w&-&+&\ul&Z_p&Z_2&Z&\pk \cr
\noalign{\srule}
L(p,q_2)&w+&-&+&\ul&Z_p&Z_2\t Z_2&Z\t Z&\pk\t\pk  \cr
\noalign{\srule}
L(p,q_3)&w&-&-&\ul&Z_p&Z_2&Z\t Z&\pk\t\pk  \cr
\noalign{\srule}
L(p,q_4)&w&-&+&\ul&Z_p&Z_2&Z\t Z&\pk\t\pk \cr
\noalign{\trule}
RP^3&+&-&-&+&Z_2&1&0&0 \cr
\noalign{\srule}
S^3&+&-&-&-&1&1&0&0 \cr
\noalign{\trule\vskip 3.0 pt\trule}
S^2\t S^1&/&-&-&+&Z&Z_2\t Z_2&Z&\pk\t \pi_k(S^2) \cr
\noalign{\trule \vskip 3.0 pt \trule}
R^3/G_1&/&+&-&+&Z\t Z\t Z&{\rm St}(3,Z)&0&\pk \cr
\noalign{\srule}
R^3/G_2&/&+&-&+&Z\t Z_2\t Z_2&\aut (G_2)&0&\pk \cr
\noalign{\srule}
R^3/G_3&/&+&+&+&Z\t Z_3&\aut (G_3)&0&\pk \cr
\noalign{\srule}
R^3/G_4&/&+&+&-&Z\t Z_2&\aut (G_4)&0&\pk \cr
\noalign{\srule}
R^3/G_5&/&+&+&+&Z&\aut (G_5)&0&\pk \cr
\noalign{\srule}
R^3/G_6&/&+&+&-&Z_4\t Z_4&\aut (G_5)&0&\pk \cr
\noalign{\trule}
S^1\times R_g&/&+&-&-&Z\t Z_{2g}&\aut (Z\t F_g)&0&\pk \cr
\noalign{\trule}
K(\pi, 1)_{\hbox{sl}}&/&+&*&*&A\pi&\aut (\pi)&0&\pk\cr
\noalign{\trule}
}}
$$
\par}
\vfill\break

\beginsection{Appendix:}

\advance\hoffset by 1.0truecm
\advance\hsize by -2truecm         
\advance\vsize by -2truecm

In this appendix we investigate in some more detail the diffeomorphisms of
the $l$-fold connected sum of handles:
$$
\S=\biguplus_{i=1}^l S^1\times S^2\,.
\eqno{(A1)}
$$
Its fundamental group is given by the $l$-fold free product of $Z$:
$$
\pi_1(\S)=\bigast_{i=1}^l Z =:F_l\,,
\eqno{(A2)}
$$
where $F_l$ just stands for the free group on $l$ generators. We visualize
the generator $g_i$ of the $i$'th $Z$ as a smooth, nonintersecting
and oriented curve which starts from some basepoint, enters the $i$'th
handle through $S_{(i,1)}$, leaves it through $S_{(i,2)}$, and returns to
the basepoint. The direction defined by $g_i$ is called positive.
By $g_ig_j$ we denote a curve that first traverses the $i$'th and
then the $j$'th handle in positive directions.

According to eq. (1.12) and the following remarks, one has [McC1]
$$
\pi_0(D_F(\S))=\left(\bigoplus_{i=1}^l Z_2\right)\times\pi_1(C_1)\,,
\eqno{(A3)}
$$
where the $i$'th $Z_2$ is generated by a belt-twist of the $i$'th handle.

As remarked earlier, $\pi_1(C_1)$ is generated by slides, exchanges, and
spins. In paragraph 4.3 of ref. [La] it is shown that the following
sequence is exact:
$$
\diag{
0&\mapright{}&\mathop{\bigoplus}_{i=1}^l Z_2&\mapright{}&
\pi_0(D^+_{\infty}(\S))&\mapright{}& Aut(F_l)&\mapright{}&1\cr}\,,
\eqno{(A4)}
$$
where $\oplus_{i=1}^l Z_2$ is the same as above (i.e. the $i$'th $Z_2$ is
generated by a belt-twist of the $i$'th handle). Now, the handle
manifold $S^1\times S^2$ is non-spinorial so that $\S$ is also
non-spinorial. But for non-spinorial 3-manifolds one has [Gi1][Wi1]
$$
\pi_0(D_F)\cong\pi_0(D^+_{\infty})\,.
\eqno{(A5)}
$$
Together (A3-5) imply
$$\eqalignno{
&\pi_0(D_F(\S)=\left(\bigoplus_{i=1}^l Z_2\right)\times Aut (F_l)\,,
&(A6)\cr
&\hbox{that is}\quad\pi_1(C_1)\cong Aut (F_l)\,.
&(A7)\cr}
$$

In quantum gravity one is e.g. interested in some of the representation
properties of $\pi_0(D_F)$ which we can now investigate. We are interested
in the question of how slides interact with the other operations, in
particular the exchanges. The reason being that slides generate those
diffeomorphisms which mix the interior and exterior of primes (as explained
in section 1) and are thus harder to interpret physically, at least in an
approximation where the primes are treated as individual particle like
entities (Geons). In view of this, one might e.g. be interested in the
following question: Under what conditions may we forget about slides? A
first simple answer for the example considered here is provided by
theorem A below.

Since internal diffeomorphisms (here the belt twists) can be given any
representation independent of the rest (product structure of (A3)) we
may restrict attention to the group $\Gamma_l:=Aut(F_l)$.
We will follow chapter 7 of ref. [CM] in our description of the
presentations for $\Gamma_l$. $\Gamma_l$ can be generated by four
generators: $P$, $O$, $Q$ and $U$, whose action on the $g_i$ is given by
$$\eqalignno{
P&:[g_1,g_2,g_3,\dots ,g_l]\mapsto [g_2,g_1,g_3,\dots ,g_l]       &(A8a)\cr
Q&:[g_1,g_2,g_3,\dots ,g_l]\mapsto [g_2,g_3,g_4,\dots ,g_1]       &(A8b)\cr
O&:[g_1,g_2,g_3,\dots ,g_l]\mapsto [g_1^{-1},g_2,g_3,\dots ,g_l]  &(A8c)\cr
U&:[g_1,g_2,g_3,\dots ,g_l]\mapsto [g_1g_2,g_2,g_3,\dots ,g_l]\,. &(A8d)\cr}
$$
Their ``physical'' interpretation is as follows: $P$ exchanges handle no 1
and handle no 2, $Q$ exchanges all $l$ handles in cyclic order, $O$ spins
handle no 1, and $U$ slides the second end of the first handle through the
second handle in a positive direction. $P$ and $Q$ alone generate the
permutation group of $l$ objects which is a sub- but no factor group
of $\Gamma_l$, and $P,Q,O$ generate an order $2^ll!$ subgroup
with obvious interpretation (it may be characterized as the group of
rearangements of $l$ books in a shelf with no orientations given to the
backs). The element $(OU)^2$ ($=(UO)^2$) represents a slide of the whole
of the first handle through handle two. If, in addition to the relations
given below, one imposes the relation $(OU)^2=E$, one obtains a
presentation of the group $GL(l,Z)$ which is a factor group of $\Gamma_l$
[CM].

We shall now give the relations for the generators $P,Q,O,U$ where $E$
denotes the identity. Given them, we then study some quotient groups by
imposing additional relations $R_i(P,Q,O,U)=E$ by hand which gives us a
presentation of $\Gamma_l/N_R$, where $N_R$ is the smallest normal subgroup
in $\Gamma_l$ containing the elements $R_i(P,Q,O,U)$. The reason for this
is that a representation $\rho:\Gamma_l\rightarrow GL(l,C)$ satisfying
$R_i(\rho(P),\rho(Q),\rho(O),\rho(U))=1$ comes from a representation of
$\Gamma_l/N_R$, which, in the cases we choose, is very simple.
We can thus imediately give all the representations of $\Gamma_l$ which
satisfy these relations.

$\Gamma_1$ is uninteresting and just given by $Z_2$, generated by the
single spin $O$. For $\Gamma_2$ one has $P=Q$ and the relations for
the remaining generators are:
$$
P^2=O^2=(PO)^4=(POPU)^2=(POU)^3=E,\quad (OU)^2=(UO)^2\,.
\eqno{(A9)}
$$
Let us look at this case first before going to the general case.

$\bullet$ {\it Abelian representations} : The presentation of the
abeliansation $A\Gamma_2$ of $\Gamma_2$ is easily obtained (all
generators commute) and given by
$$
P^2=O^2=U^2=POU=E
\eqno{(A10)}
$$
which is just the group $Z_2\times Z_2$ generated by $P$ (left $Z_2$)
and $O$ (right $Z_2$) and where $U$ generates the diagonal $Z_2$. For
an abelian representation this implies that any of the genrators $P,O,U$
is non-trivially represented precisely if one of the others is. There is no
$P-Q$ correlation unless one imposes $U=E$. This is also true generally
since the factor group $U=E$ is abelian as we show below. One easily checks
from (A9) that taking any of the generators $P$, $O$ or $U$ to commute
with the other two already implies commutativity of all generators.
Moreover, it is in fact sufficient to require exchanges to commute with
slides only. Proof: If $P$ and $U$ commute $(POPU)^2=P(OU)^2P=E\Rightarrow
(OU)^2=E$ and $(POU)^3=POP(UO)^2PU=POU=E\Rightarrow PU=O$, so that $P$ also
commutes with $O$.

$\bullet$ {\it Slides represented trivially} : Setting $U=E$ yields
$$
P^2=O^2=(PO)=E
\eqno{(A11)}
$$
which is just $Z_2$ generated by $P=O$. Thus representing $U$ trivially
leads to an abelian representation with $P-O$ correlation.

$\bullet$ {\it $P-O$ correlating representations} : Setting $P=O$ yields
$$
P^2=U^3=(PU)^2=E
\eqno{(A12)}
$$
which is the presentation of the permutation group $S_3$ of 3 objects
(e.g. $P=(12)$ and $U=(123)$) or, equivalently, of $D_6$, the
dihedral group of order 6 which describes the symmetries of an
equilateral triangle ($P$ generates reflections about a symmetry axis,
$U$ a rotation by $2\pi/3$). There are two non-trivial representations,
the one-dimensional one, $R_1$, and the two dimensional one, $R_2$:
$$\eqalign{
R_1&:\qquad P\mapsto -1,\quad U\mapsto 1                         \cr
R_2&:\qquad P\mapsto\pmatrix{-1&0\cr 0&1\cr},\quad
            U\mapsto\pmatrix{-{1\over 2}&{\sqrt{3}\over 2}\cr
                           -{\sqrt{3}\over 2}& -{1\over 2}\cr}\,.\cr}
\eqno{(A13)}
$$
$R_2$ shows that there are $P-O$ correlating representations which also
represent $U$ non-trivially.

$\bullet${\it Spins represented trivially} : Setting $O=E$ yields
$$
P^2=U^2=(PU)^3=E
\eqno{(A14)}
$$
which reduces to (A12) by replacing $U\rightarrow PU$. Correspondingly, the
two non-trivial representations follow from (A14). This shows how slides
can interact with exchanges without involving spins.

Next we turn to the general case $l>2$. A complete list of relations is
given below (the relations are not independent).
$A\leftrightarrow B$ means that $A$ and $B$ commute.
$$\eqalignno{
&P^2=E                                                   &(A15a)\cr
&(QP)^{n-1}=Q^n                                          &(A15b)\cr
&P\lr Q^{-i}PQ^i\qquad\hbox{for}\ 2\leq i\leq{l\over 2}  &(A15c)\cr
&O^2=E                                                   &(A15d)\cr
&O\lr Q^{-1}PQ                                           &(A15e)\cr
&O\lr QP                                                 &(A15f)\cr
&(PO)^4=E                                                &(A15g)\cr
&U\lr Q^{-2}PQ^2\qquad\hbox{for}\ l>3                    &(A15h)\cr
&U\lr QPQ^{-1}PQ                                         &(A15i)\cr
&U\lr Q^{-2}OQ^2                                         &(A15j)\cr
&U\lr Q^{-2}UQ^2\qquad\hbox{for}\ l>3                    &(A15k)\cr
&U\lr UOU                                                &(A15l)\cr
&U\lr PQ^{-1}OUOQP                                       &(A15m)\cr
&U\lr PQ^{-1}PQPUPQ^{-1}PQP                              &(A15n)\cr
&(PQ^{-1}UQ)^2=UQ^{-1}UQU^{-1}                           &(A15o)\cr
&U^{-1}PUPOUOPO=E                                        &(A15p)\cr
&(POPU)^2=E                                              &(A15q)\cr}
$$

$\bullet${\it Abelian representations} :
In this case (A15) boils down to the following relations for commuting
generators:
$$
P^2=O^2=PO=P^{l-1}Q^{-1}=U=E\,,
\eqno{(A16)}
$$
which is the presentation of $Z_2$ generated by $P$ ($= O$). $Q$ equals $E$
for $l$ odd and $P$ for $l$ even. The difference of this case to $l=2$ lies
in the condition $U=E$ which followed from (o) in view of (a). In fact, (o)
implies $U=E$ already from the commutativity of $U$ with $P$ and $Q$.
Abelian representations (equivalently, representations for which slides and
exchanges commute) thus necessarily represent slides trivially.

$\bullet$ {\it Slides represented trivially} : If we set $U=E$, (p) and (a)
imply $P=O$, (f) says that $P$ and $Q$ commute, and (b) then implies that
$P^{l-1}=Q$, i.e. that $Q=E$ for $l$ odd and $Q=P$ for $l$ even. We thus
obtain the group $Z_2$ generated by $P$ ($=O$).

$\bullet$ {\it $P-O$ correlating representations} : If we impose $P=O$,
(f) implies $P\leftrightarrow Q$, (i) implies $U\leftrightarrow Q$ and
(j) implies $U\leftrightarrow P$ so that all generators commute.

We can summarize these points in the following

\proclaim Theorem A. Let $\rho$ be a representation for $\Gamma_l$,
$l\geq 3$. The following conditions on $\rho$ are equivalent:
$$\eqalign{
  &a:\,\hbox{slides and exchanges commute}                          \cr
  &b:\,\hbox{$\rho$ is abelian}                                     \cr
  &c:\,\hbox{slides are represented trivially}                      \cr
  &d:\,\hbox{$\rho$ correlates $P$ and $O$, i.e. $\rho(P)=\rho(O)$}\,
                                                          \bullet \cr}
$$

\vskip 2.0truecm

\centerline{\bf ACKNOWLEDGMENTS}

I thank D. McCullough for pointing out ref. [LSY],
and K.B. Lee for providing me with a copy of this reference.

\vfill\eject

\beginsection{References}

\advance\hsize by -1.3truecm
\advance\hoffset by 1.3truecm

\item{[ABBJRS]}  Aneziris, C., Balachandran, A.P., Bourdeau, M., Jo, S.,
                 Ramadas, T.R., Sorkin, R.: Aspects of spin and statistics
                 in generally covariant theories. Int. Jour. Mod. Phys. A,
                 {\bf 4}, 5459-5510 (1989)

\item{[FS]}      Friedman, J., Sorkin, R.: Spin ${1\over 2}$ from Gravity.
                 Phys. Rev. Lett., {\bf 44}, 1100-1103 (1980).

\item{[GH]}      Gibbons, G.W., Hawking S.W.: Selection rules for topology
                 change. Commun. Math. Phys. {\bf 148}, 345-352 (1992)

\item{[Gi1]}     Giulini, D.: On the configuration space topology in
                 General Relativity. Preprint, Freiburg THEP-92/32 and
                 gr-qc 9301020. Submitted for publication, August 1993.

\item{[Gi2]}     Giulini, D.: Quantum mechanics on spaces with finite
                 non-abelian fundamantal group. In preparation.

\item{[Gi3]}     Giulini, D.: On the selection rules for spin-Lorentz
                 cobordisms. Commun. Math. Phys. {\bf 148}, 353-357 (1992).

\item{[GiL]}     Giulini, D., Louko, J.: No-boundary $\theta$-sectors in
                 spatially flat quantum cosmology. Phys. Rev. D, {\bf 46},
                 4355-4364 (1992).

\item{[GrL]}     Gromov, M., Lawson, B.: positive scalar curvature and the
                 Dirac operator on complete Riemannian manifolds. Institut
                 des Hautes \'Etudes Scientifique (I.H.E.S.), publicationes
                 mathematiques {\bf 58}, 295-408 (1983).

\item{[He]}      Hempel, J.: 3-Manifolds. Annals of mathematical studies,
                 Vol. 86, Princeton University Press (1976).

\item{[HL]}      Hendriks, H., Laudenbach, F.: Diffeomorphismes des sommes
                 connexes en dimension trios. Topology, {\bf 23}, 423-443
                 (1984).

\item{[HM]}      Hendriks, H., McCullough, D.: On the diffeomorphism group
                 of a reducible 3-manifold. Topology and its Application,
                 {\bf 26}, 25-31 (1987).

\item{[HW]}      Hartle, J., Witt, D.: Gravitational $\theta$-states and
                 the wave function of the universe. Phys. Rev. D, {\bf 37},
                 2833-2836 (1988).

\item{[Is]}      Isham, C.J.: Topological $\theta$-sectors in canonically
                 quantized gravity. Phys. Lett. B, {\bf 106}, 188-192
                 (1981).

\item{[La]}      Laudenbach, F.: Topologie de la dimension trois; homotopie
                 et isotopie. Asterisque, {\bf 12}, 1-137 (1974).

\item{[LSY]}     Lee, K.B., Shin, J., Yokura, S.: Free actions of finite
                 abelian groups on the 3-torus. University of Oklahoma
                 preprint, 1993.

\item{[McC1]}    McCullough, D.: Mappings of reducible manifolds.
                 Geometric and Algebraic Topology; Banach center
                 publications, {\bf 18}, 61-76 (1986). PWN - Polish
                 scientific publishers, Warszawa.

\item{[McC2]}    McCullough, D.: Private communication.

\item{[Mi] }     Milnor, J.: Introduction to algebraic K-theory. Annals of
                 Mathematics Studies Vol. 72, Princeton University Press
                 (1971)

\item{[O]}       Orlik, P.:  Seifert manifolds. Lecture notes in
                 mathematics Vol. 291, Springer Verlag, Berlin, Heidelberg,
                 New York 1972.

\item{[RS]}      Rourke, C. de S\'a, C.; The homotopy type of
                 homeomorphisms of 3-manifolds. Bull. Amer. Math. Soc.,
                 {\bf 1}, 251-254 (1974).

\item{[So]}      Sorkin, R.: Classical topology and quantum phases: Quantum
                 Geons. In: Geometrical and algebraic aspects of nonlinear
                 field theory; Ed. De Filippo, S., Marinaro, M., Marmo, G.
                 and Vilasi, G.; Elsevier Science Publishers B.V. (North
                 Holland) (1989).

\item{[Th]}      Thomas, C.B.: Elliptic Structures on 3-Manifolds. London
                 Math. Soc. Lecture Note Series 103, Cambridge University
                 Press (1986); also: Bull. London Math. Soc. {\bf 20} 65-67
                 (1988)

\item{[Wi1]}     Witt, D.: Symmetry groups of state vectors in canonical
                 quantum gravity. Jour. Math. Phys., {\bf 27}, 573-592
                 (1986).

\item{[Wi2]}     Witt, D.: Topological obstructions to maximal slices.
                 Santa Barbara preprint, UCSB-1987.
\bye